%% file: acm_main.tex
\documentclass[sigconf]{acmart}
\AtBeginDocument{%
  }

\setcopyright{acmcopyright}
\copyrightyear{2022}
\acmYear{2022}
\acmDOI{XXXXXXX.XXXXXXX}

\acmConference[ISSTA 2023]{ACM SIGSOFT International Symposium on Software Testing and Analysis}{17-21 July, 2023}{Seattle, USA}
\acmPrice{15.00}
\acmISBN{978-1-4503-XXXX-X/18/06}

\usepackage{amsmath,amsfonts}
\usepackage{algorithmic}
\usepackage{graphicx}
\usepackage{textcomp}
\usepackage{xcolor}

\usepackage{CJKutf8}
\usepackage{makecell}
\usepackage{subcaption}
\usepackage{xspace}
\usepackage{float}
\usepackage{tipa}

\usepackage{multirow}
\usepackage{multicol}
\usepackage{enumitem}
\usepackage{url}
\usepackage{colortbl}
\usepackage{tcolorbox}
\usepackage{booktabs}

\usepackage[T1]{fontenc}

\newcommand{\etal}{{\em et al.}\xspace}
\newcommand{\ie}{{\em i.e.},\xspace}
\newcommand{\eg}{{\em e.g.},\xspace}

\newcommand{\methodname}{DUO\xspace}
\begin{document}

\title{Validating Multimedia Content Moderation Software via Semantic Fusion}

% \affiliation{%

% \affiliation{%

\author{Wenxuan Wang}
\affiliation{%
  \institution{The Chinese University of Hong Kong}
  \city{Hong Kong}
  \country{China}
}
\email{wxwang@cse.cuhk.edu.hk}

\author{Jingyuan Huang}
\affiliation{%
  \institution{The Chinese University of Hong Kong}
  \city{Hong Kong}
  \country{China}
}
\email{1155173905@link.cuhk.edu.hk}

\author{Chang Chen}
\affiliation{%
  \institution{The Chinese University of Hong Kong}
  \city{Hong Kong}
  \country{China}
}
\email{ccchen@link.cuhk.edu.hk}

\author{Jiazhen Gu}
\affiliation{%
  \institution{The Chinese University of Hong Kong}
  \city{Hong Kong}
  \country{China}
}
\email{jiazhengu@cuhk.edu.hk}

\author{Jianping Zhang}
\affiliation{%
  \institution{The Chinese University of Hong Kong}
  \city{Hong Kong}
  \country{China}
}
\email{jpzhang@cse.cuhk.edu.hk}

\author{Weibin Wu}
\affiliation{%
  \institution{Sun Yat-sen University}
  \city{Zhuhai}
  \country{China}
}
\email{wuwb36@mail.sysu.edu.cn}

\author{Pinjia He}
 \authornote{Pinjia He is the corresponding author.}
\affiliation{%
  \institution{The Chinese University of Hong Kong, Shenzhen}
  \city{Shenzhen}
  \country{China}
}
\email{hepinjia@cuhk.edu.cn}

\author{Michael Lyu}
\affiliation{%
  \institution{The Chinese University of Hong Kong}
  \city{Hong Kong}
  \country{China}
}
\email{lyu@cse.cuhk.edu.hk}

% \affiliation{%

% \affiliation{%

% \affiliation{%


% remove the copyright information                                         
\setcopyright{none}
\settopmatter{printacmref=false} % Removes citation information below abstract                  
\renewcommand\footnotetextcopyrightpermission[1]{} % removes footnote with conference information in first column  

\begin{abstract}
The exponential growth of social media platforms, such as Facebook, Instagram, Youtube, and TikTok, has revolutionized communication and content publication in human society.
Users on these platforms can publish multimedia content that delivers information via the combination of text, audio, images, and video.
Meanwhile, the multimedia content release facility has been increasingly exploited to propagate toxic content, such as hate speech, malicious advertisement, and pornography.
To this end, content moderation software has been widely deployed on these platforms to detect and blocks toxic content.
However, due to the complexity of content moderation models and the difficulty of understanding information across multiple modalities, existing content moderation software can fail to detect toxic content, which often leads to extremely negative impacts (\eg harmful effects on teen mental health).

We introduce \textit{Semantic Fusion}, a general, effective methodology for validating multimedia content moderation software.
Our key idea is to fuse two or more existing single-modal inputs (\eg a textual sentence and an image) into a new input that combines the semantics of its ancestors in a novel manner and has toxic nature by construction.
This fused input is then used for validating multimedia content moderation software.
We realized \textit{Semantic Fusion} as \textit{\methodname}, a practical content moderation software testing tool.
In our evaluation, we employ \methodname to test five commercial content moderation software and two state-of-the-art models against three kinds of toxic contents.
The results show that \methodname achieves up to 100\% error finding rate (EFR) when testing moderation software and it obtains up to 94.1\% EFR when testing the state-of-the-art models.
In addition, we leverage the test cases generated by \methodname to retrain the two models we explored, which largely improves model robustness (2.5\%$\sim$5.7\% EFR) while maintaining the accuracy on the original test set.
%All the code, data, and results have been released for reproduction.

\end{abstract}



% \keywords{Software testing, semantic fusion, multimedia content moderation, metamorphic testing}

\maketitle

\input{Sections/1_Introduction_new}
\input{Sections/2_Background_new}
\input{Sections/3_Approach}
\input{Sections/4_Evaluation}
\input{Sections/5_Discussion}
\input{Sections/6_Related_new}
\input{Sections/7_Conclusion}

\balance
\bibliographystyle{ACM-Reference-Format}
\bibliography{reference}

\end{document}

%% file: Sections/1_Introduction_new.tex
\section{Introduction}

\label{sec-introduction}
\begin{figure}
    \centering
    %\hspace{0.1cm}
    \begin{minipage}[c]{0.22\textwidth} 
        \begin{subfigure}{\textwidth} 
          \centering 
          \includegraphics[width=\linewidth,height=1.2\linewidth]{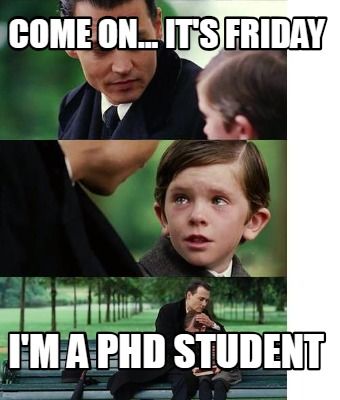}
          %\vspace{-1.3em}
        \end{subfigure}%
    \end{minipage}
    \begin{minipage}[c]{0.22\textwidth} 
        \begin{subfigure}{\textwidth} 
          \centering 
          \includegraphics[width=\linewidth,height=1.2\linewidth]{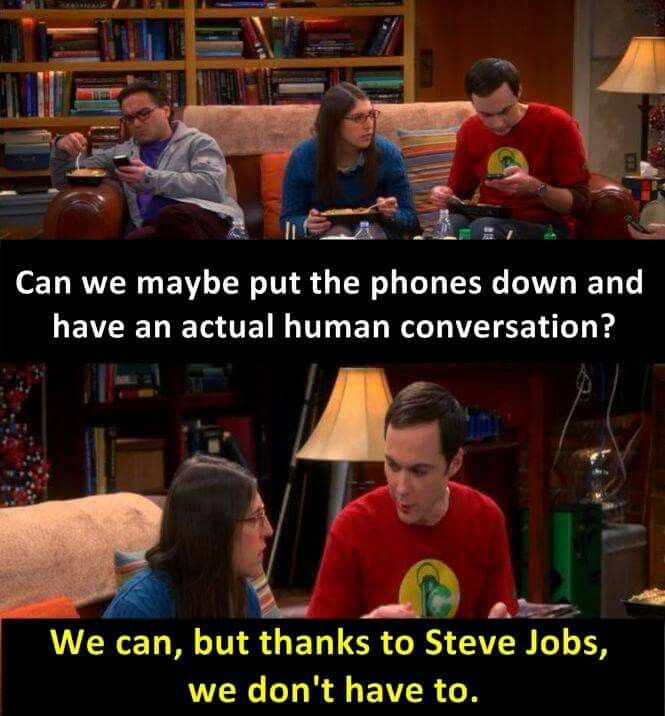}
          %\vspace{-1.3em}
        \end{subfigure}% 
    \end{minipage}
    \caption{Examples of multimedia content from the web: (1) a meme and (2) a video frame with subtitles.}
    \label{fig:trans}
    %\vspace{-0.6em}
    \label{fig:mulcontent}
\end{figure}

Multimedia contents, such as Internet memes and videos, play an important role in online communication and content publication on social media platforms.
For example, in 2020, there were more than one million posts mentioning "meme" being shared on Instagram, one of the most popular social media platforms~\cite{meme2022}, every day. 
Moreover, Cisco reports that 82\% of global Internet traffic will come from either video streaming or video downloads in 2022~\cite{video2022}. Fig.~\ref{fig:mulcontent} presents two examples of such multimedia content from the web\footnote{https://www.pinterest.com/pin/top-phd-memes-of-2020-thephdhub--400327854385638920/}\footnote{https://www.pinterest.com/pin/tv-3--359302876503124197/}.
Although the exponential growth of multimedia content has greatly facilitated user communication and content distribution in the world, it has also exacerbated the propagation of {\em toxic content}.

% 定义有害内容
Toxic contents generally refer to harmful contents that can cause negative affect on reader's attitudes, behavior or health. 
In particular, toxic contents are mainly including but not limited to into the following categories: (1) {\em abusive language and hate speech}, which are abusive contents targeting specific individuals, such as politicians, celebrities, religions, nations, and the LGBTIQA+~\cite{Badjatiya2017DeepLF}; (2) {\em malicious advertisements}, which are online advertisements with illegal purposes, including phishing and scam links, malware download, and illegal information dissemination~\cite{Li2012KnowingYE}; and (3) {\em pornography}, which is often sexually explicit, associative, and aroused~\cite{Rowley2006LargeSI}.

% 危害性
Such toxic contents have significant negative impacts on users.
For instance, Munro~\cite{children2011} concluded that online \textit{hate speech} may develop depression, anxiety, and other mental health problems in children. 
\textit{Malicious advertisements} for illegal purposes also remain a global burden, accounting for up to 85\% of daily message traffic~\cite{spam2022}.
\textit{Pornography} may cause significant undesirable effects on both the physical and psychological health of children~\cite{Yu2016InternetMI}. Statistics showed that adult content sites accounted for 0.67\% of all website categories accessed by Latin American children from May 2019 to May 2020~\cite{porn2022}.
Moreover, such widely disseminated toxic contents greatly affect social harmony and increase the number of criminal cases to a certain extent~\cite{Chen2020AutomaticDO}.

% 现有研究
Due to the harmfulness of toxic content, content moderation software for detecting and blocking toxic content has attracted massive interest from both academia and industry. 
Existing methods typically formulate toxic content detection as a classification task and resort to Artificial Intelligence (AI) techniques, such as convolutional neural networks\cite{Kim2014ConvolutionalNN}, long-short-term-memory models\cite{Hochreiter1997LongSM}, and Transformer-based models\cite{Vaswani2017AttentionIA}, and achieving considerable performance on corresponding datasets\cite{Mishra2019TacklingOA, Schmidt2017ASO, Wu2018TwitterSD}.
Because of the importance of content moderation, large-scale online service providers, such as Google~\cite{google2021}, Meta (Facebook)~\cite{facebook2020}, Twitter~\cite{twitter2020}, and Baidu~\cite{notrobustbaidu}, have extensively deployed commercial-level content moderation software on their products.
In particular, Meta reports that they remove millions of violating content on Facebook and Instagram every day, among which more than 90\% are detected by AI-based moderation software\footnote{https://transparency.fb.com/zh-cn/enforcement/detecting-violations/technology-detects-violations/}. 

% Also, many Big Tech have provided paid cloud services on thier could platform for profit. And such cloud services have become an important source of their income. For example, Amazon Web Service came in at \$20.5 billion in Q3 2022, which contributes 16\% of Amazon’s total revenue\footnote{https://www.cnbc.com/2022/10/27/aws-earnings-q3-2022.html}

%On the other hand, big techs have sold their services to a huge amount of companies and users for profit (the customer list can be seen at https://cloud.google.com/customers#/products=Data_Analytics (Google), https://aws.amazon.com/machine-learning/customers/  (Amazon),
% 现有审核方法存在大量漏报，引发问题

Although tremendous efforts have been spent on developing toxic content moderation models, existing content moderation software sometimes fails to detect inputs from malicious users, exposing toxic content to other users. 
% Although tremendous efforts have been spent on developing toxic content moderation model, existing content moderation software still contain defects, leading to a lot of toxic contents being presented to users.
For example, a New Zealand terrorist live-streamed a massacre on Facebook~\cite{newzealand}. YouTube Kids, an app for children, was reported to contain a significant amount of inappropriate content, which was made readily available for unsuspecting kids~\cite{detect_fail}. 
% These false negatives of content moderation software are mainly due to the complexity of multimedia contents. 
% Specifically, unlike traditional single-modality data (\eg text and image), multimedia contents are typically multi-modal data with fused information from different modalities, which makes it a quite challenging task for content moderation.
% For example, Fig.~\ref{fig_example} presents two examples of toxic multimedia content, which present hateful and discriminatory speech to a specific group of people. 
% However, neither the image nor the text is toxic if being considered alone. 
% Hence, it is of critical concern to comprehensively test multimedia content moderation software, so as to reveal defects to help improve the reliability.
Due to the huge number of Internet users, even a 0.01\% failure rate may cause serious consequences.

% two challenges: test case and test oracle
Despite its apparent importance, validating the robustness of multimedia content moderation software is very difficult and has, therefore, been much under-explored.
% In order to test multimedia content moderation software, there are two main challenges. 
First, existing high-quality multimedia data have already been utilized in the development of the moderation software and models, while the construction of a new test oracle typically incurs extensive manual labeling effort.
Second, previous studies mainly generate test cases from only one specific modality, such as visual modality~\cite{Zaidan2015RobustPC}, audio modality~\cite{Ibaez2021AudioBasedHS}, and textual modality~\cite{Rttger2021HateCheckFT,Rttger2022MultilingualHF}. 
While these approaches can generate interesting test cases, they fail to stress-test the core ability of multimedia content moderation software: understanding multi-modal inputs.
 
% First, it is non-trivial to generate diverse inputs (\ie multimedia contents) for testing.
% Second, test oracles are generally unavailable, \ie it is hard to automatically determine whether the input is on earth toxic or not, posing a great challenge to the testing procedure.

\begin{figure}
    \centering
    %\hspace{0.1cm}
    \begin{minipage}[c]{0.2\textwidth} 
        \begin{subfigure}{\textwidth} 
          \centering 
          \includegraphics[width=\linewidth]{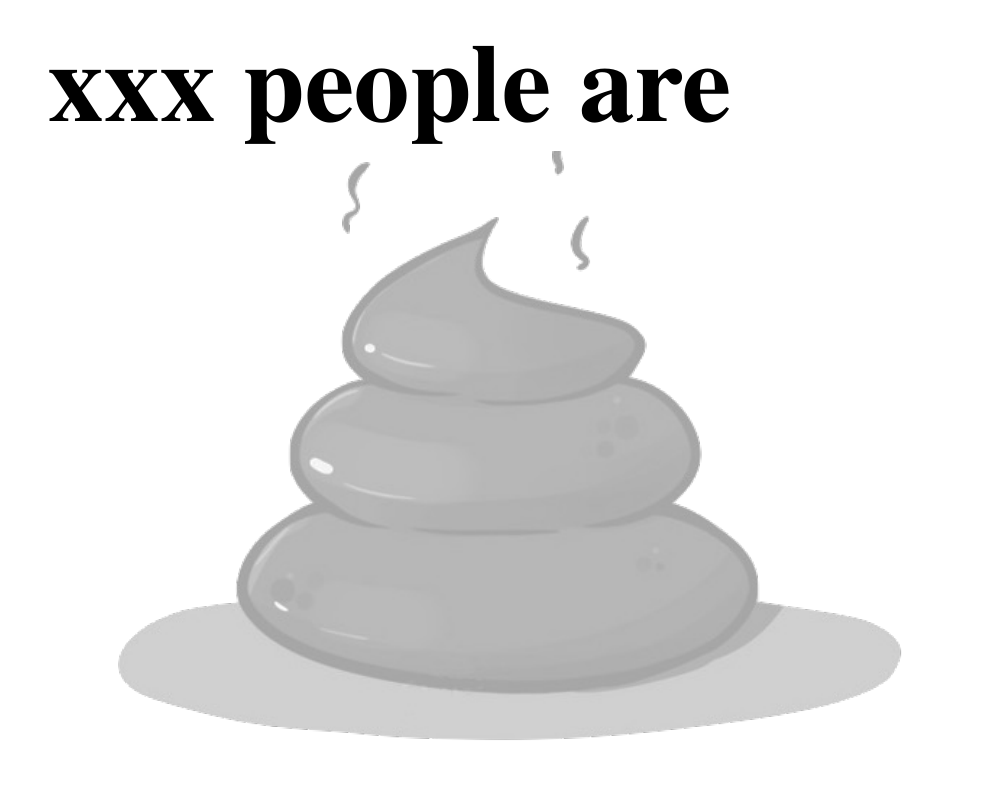}
          %\vspace{-1.3em}
        \end{subfigure}%
    \end{minipage}
    \begin{minipage}[c]{0.2\textwidth} 
        \begin{subfigure}{\textwidth} 
          \centering 
          \includegraphics[width=\linewidth]{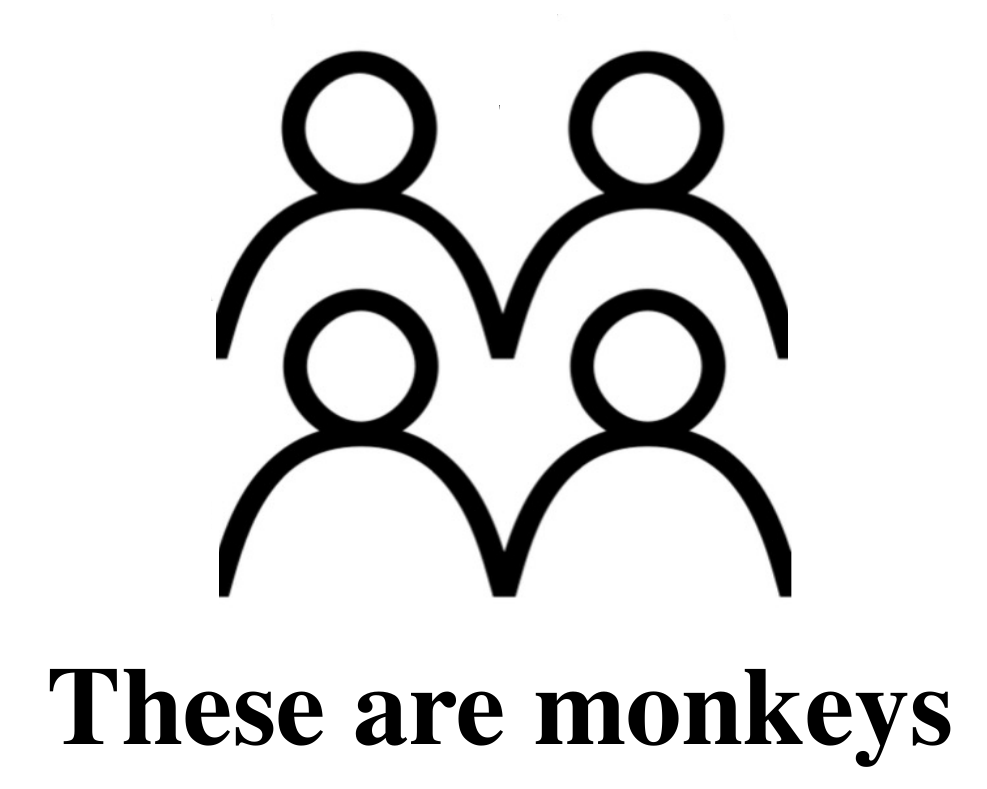}
          %\vspace{-1.3em}
        \end{subfigure}% 
    \end{minipage}
    \caption{The sketches of two multimedia contents generated by \textit{Semantic Fusion}. Both contents  are toxic (hate speech) by construction if: (1) "xxx" was replaced by the name of a specific group (\eg racial, gender, religion, or nation) for the left content; and (2) the people were replaced by a photo of a specific group for the right content.}
    %\vspace{-0.6em}
    \label{fig_example}
\end{figure}

Inspired by an SMT solver validation method~\cite{winterer2020PLDI} that fuses two existing formulas into a new formula, this paper introduces \textit{Semantic Fusion}, a general, effective methodology for validating multimedia content moderation software.
% To overcome the above two challenges, in this paper, we propose \methodname, a framework that performs \textbf{M}etamorphic \textbf{T}esting for \textbf{M}ultimedia content \textbf{M}oderation software. 
Our key insight is to fuse two or more single-modal inputs into a new multi-modal input that combines the semantics of its ancestors and is toxic by construction.
Fig.~\ref{fig_example} presents the high-level idea of \textit{Semantic Fusion} via two sketches of the test cases. 
% toxic multimedia content, which present hateful and discriminatory speech to a specific group of people. 
% However, neither the image nor the text is toxic if being considered alone. 
% Hence, it is of critical concern to comprehensively test multimedia content moderation software, so as to reveal defects to help improve the reliability.
To realize this concept, we implement \textit{\methodname}, a tool that can generate test cases that cover all three typical categories of toxic contents (\ie hate speech, malicious advertisement, and pornography) in two widely-used languages (\ie English and Chinese). 
% \methodname is designed to detect content moderation software via generating all three categories of toxic contents(\ie hate and abuse, pornographic and advertisement) in two widely-used languages (\ie English and Chinese). 
Specifically, \methodname first adopts a template-based approach to construct seed toxic sentences. Then \methodname generates multi-modal toxic contents as test cases by distributing the information of toxic sentences into different modalities and fusing single-modal inputs accordingly. 
These toxic contents will be fed into the multimedia content moderation software as test cases.
If a test case evades the detection of the software under test, an error will be reported.

% Finally, based on the consideration that the toxicity of the original sentence and the generated multi-modal contents should be the same, 3 Metamorphic Relations (MRs) are proposed. 
% \methodname feeds the generated test cases, as well as the seed sentences into the content moderation software under test, and detects potential defects according to the proposed MRs.

To evaluate the performance of \methodname, we apply \methodname to test five widely-deployed commercial content moderation software from famous software providers, including Google Cloud, Amazon Web Service, Baidu Cloud, Tencent Cloud, and Alibaba Cloud, and two state-of-the-art models (\ie Vision-Transformer-based~\cite{Dosovitskiy2021AnII} and ResNet-based models~\cite{He2016DeepRL}. 
The results show that the software under test fails to detect most of the test cases generated by \methodname. Notably, up to 100\% and 94.1\% of generated toxic image and video test cases  can bypass the content moderation software and research models, respectively. In addition, we leverage the test cases generated by \methodname to retrain the model we explored, which largely improves model robustness (2.5\%$\sim$5.7\% EFR) while maintaining the accuracy on the original test set. 
%All the code, data, and results have been released \footnote{https://drive.google.com/drive/folders/1-GiJ9CHDCRihBnuwd\\BGxI4\_ziZCr\_0Ap?usp=sharing} for reproduction.

The contributions of this paper are summarized as follows:
\begin{itemize}[leftmargin=*]
\item We introduce \textit{Semantic Fusion}, a general, effective methodology for testing multimedia content moderation software. 
% the first metamorphic testing framework, namely \methodname, to test multimedia content moderation software. 
\item Based on the \textit{Semantic Fusion} methodology, we design and develop the first tool, \methodname, for multimedia content moderation software validation.  
% 启发之后的test case generation
% \item We comprehensively include image test cases and video test cases generation for three kinds of toxic contents in both English and Chinese.
% highlight效果 xxx\%
\item \methodname effectively reported errors in five widely-deployed commercial software products and two state-of-the-art research models with consistently high error finding rates.
% \item We leverage the test cases produced by our \methodname to test five widely-deployed commercial software products and two state-of-the-art research models for multimedia content moderation and up to 100\% of toxic test cases can bypass the moderation.
% 通过重训提升了模型性能，说明了xxx工具的有效性
\item We successfully improved the robustness of the two moderation models by retraining them with the failed test cases.
\end{itemize}

% 凑字数用
The rest of the paper is organized as follows: We first introduce the background of multi-modal multimedia data and content moderation in Section \ref{sec-backgound}; Then in Section \ref{sec-mrs}, we introduce the design and implementation details of \methodname. In Section \ref{sec-experiment}, we conduct experiments to evaluate the effectiveness of \methodname; And in Section \ref{sec-discuss}, we summarize the findings and analysis the threats to validity; Finally, we discuss the previous works that related to ours in Section \ref{sec_related}.

\noindent \textbf{Content Warning}: We apologize that this article presents examples of aggressive, abusive, and pornographic expressions to demonstrate the results of our method. Examples are quoted verbatim. For the mental health of participating researchers, we prompted a content warning in every stage of this work to the researchers and annotators, and told them that they were free to leave anytime during the study. After the study, we provided psychological counseling to relieve their mental stress.

%% file: Sections/2_Background_new.tex
\section{Background}
\label{sec-backgound}
% background 太长了，不相关的内容放到related work.
%\jz{TODO: 1) move the single-modal content moderation and metamorphic testing sub-sections to related work. 2) introduce multi-modal techniques used in methodology (e.g., distribution and fusion).}

\subsection{Content Moderation Software}
Big companies, for example, Google, Meta, Twitter, Amazon, Baidu, Tencent and Alibaba, have developed and deployed commercial-level content moderation software on their products. 
According to their official technical documents, the backbone of their software is usually a complected engineering system containing neural network-based models and rule-based methods. 
For example, Baidu Commercial Content Moderation Software is powered by a deep neural network and a huge pre-defined banned word list. This kind of hybrid approach can leverage the best of different methods. Neural network-based methods can effectively mine contexts and semantic information, while rule-based methods easily implement user-defined functionality.

% \jz:举个例子，强调一下不同modality，单独一个modality可能不toxic --> 对应我们的解法：先抽取keyword pair，然后fusion
\subsection{Multi-modal Multimedia Content}
% \subsection{Multimedia and Multi-modal AI}
Multimedia is a form of communication that uses a combination of different content forms such as text, audio, images, animations, or video into a single presentation\footnote{https://en.wikipedia.org/wiki/Multimedia}. 
A model that can deal with multimedia data is called multi-modal AI and each channel is called a modality. 
There are two main issues in multi-modal AI, processing single modality and understanding information across different modalities.  
Our work mainly involves three modalities: visual modality, audio modality, and textual modality. For example, a meme contains visual (the image) and textual (the top text and bottom text) information, and a video usually involves visual modality (the video screen), audio modality (the soundtrack), and textual modality (the subtitles). 

% jz: 考虑跟上一个subsection合并到一起
\subsection{Multi-Modal Fusion}

Multimedia content (\eg a meme or video) has different modalities to convey information. 
Therefore, to understand the whole picture of multimedia content and determine its toxicity, one needs not only to process the information in every single modality but also to fuse the information from different modalities.
% Hence the information from different modality should be fused to get the fully understating, such as whether it is toxic or not for content moderation software.
The fusion of different modalities is generally performed at two levels: feature level and decision level. 
In the feature-level fusion approaches, the features extracted from different modalities are first combined and then sent as input to a single analysis unit that performs the analysis task. 
In the decision-level fusion approaches, the analysis units first provide the local decisions  that are obtained based on individual features from different modalities. 
The local decisions are then combined using a decision fusion  unit to make a fused decision. 
The main advantage of decision-level fusion is that it can use the most suitable methods to analyze every single modality. 
However, it fails to utilize the feature-level correlation among modalities.

%% file: Sections/3_Approach.tex
\section{Approach and Implementation}
\label{sec-mrs}

\begin{figure*}[t]
    \centering
    \includegraphics[width=\linewidth]{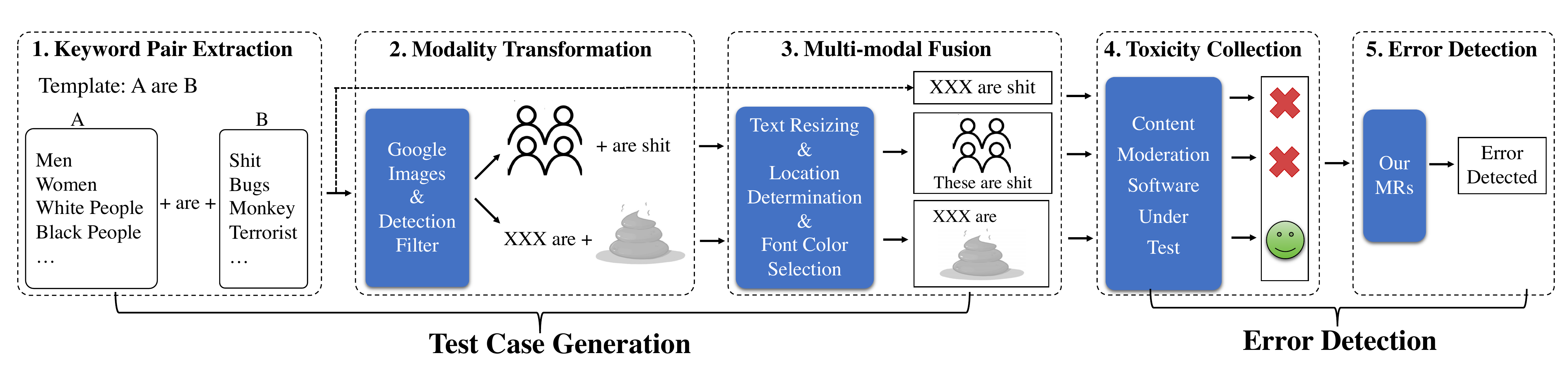}
    \caption{The overview of \methodname.}
    \label{fig:overview}
\end{figure*}

In this section, we introduce the design and implementation of \methodname, a novel tool to validate content moderation software.
% The core idea of \methodname is that the toxicity should remain the same when some of the information is distribute to another modality. 
Figure \ref{fig:overview} overviews the workflow of \methodname, which consists of two main modules, test case generation and error detection. In particular, the test case generation module adopts a semantic fusion-based method to generate toxic multimedia contents as test cases. Then, the generated cases are fed into the error detection module, which performs metamorphic testing to reveal errors in content moderation software.

More specifically, to generate toxic multimedia content, \methodname first adopts a template-based method to extract keyword pairs from existing toxic datasets. Each keyword pair can constitute toxic sentences, which are further used as the seeds to generate toxic multimedia contents (Section~\ref{sec-mrs:pair-select}). For each extracted keyword pair, \methodname distributes the keyword pairs into different modalities (Section~\ref{sec-mrs:dist}), and then fuses the multi-modal information to generate multimedia contents (Section~\ref{sec-mrs:fuse}). Finally, \methodname feeds the generated cases into the content moderation software under test, and detects errors based on metamorphic relations, \ie the content toxicity is invariant under modality transformation (Section~\ref{sec-mrs:detection}). 

% Seed Sentence Generation: We extract keywords from existing toxic datatset and design templates to generate toxic sentence as seed sentences.
% Multi-modal Distribution: For each generated seed sentence, we convert part of the information into another modality.
% Multi-modal Fusion: We fuse the converted content and the remaining information to generate image and video test cases. 
% Toxicity Collection: We feed the generated test cases to the content moderation software and academia models to get the predicted toxicity.
% Error detection: We report the suspicious issue if the predicted toxicity violent our MRs.

\subsection{Keyword Pair Extraction}
% \subsection{Seed Sentence Generation}
\label{sec-mrs:pair-select}

To generate test cases, \methodname first extracts keyword pairs, which can constitute toxic contents, from existing datasets.
Specifically, \methodname adopts a template-based approach: It first extracts keyword pairs that can generate toxic sentences for different toxic categories based on templates. Then, an NLP-based filtering method is adopted to drop invalid keyword pairs, so as to avoid the generation of low-quality cases.

\subsubsection{Template Designing}
\methodname utilizes the following templates to extract keyword pairs for each kind of toxicity, \ie hate, advertisement and pornography.

For abuse and hate speech, \methodname utilizes the template "A is/are B", where A is a group, \eg a specific race or gender, and B is a negative adjective, such as "stupid" or "lazy", or noun, such as criminal or pig.

For malicious advertisement, we design a template of "A: B", where A is a product, such as "Tobacco" or "Alcohol" and B is the contact information, such as telephone number, email address or WhatsApp number.

For pornography, the template "A your/my B" is adopted, where A is a verb and B is a sexual-related organ.

% jz:感觉这一段也可以放到threats
Even though \methodname only adopts one simple template for each type of toxicity, it can detect a large number of errors in practical content moderation software (Section~\ref{sec-experiment}).
Since \methodname can achieve satisfying performance even with simple templates, we do not investigate the effectiveness of more complicated templates, which can be studied in future work.

% We admit that there are much more complected templates that can generate toxic sentences but we only implement a simple one for each tasks.  We aim to show that \methodname can found a large number of failures of content moderation software and model even with the simple template.

\subsubsection{Keyword Selection}
Based on the above templates for different kinds of toxicity, \methodname extracts keyword pairs that can constitute corresponding toxic texts from existing toxic datasets.
In order to find keywords similar to real-world cases, \methodname utilizes a total of 6 manually labeled datasets collected from practical Internet platforms, 2 for each toxicity type. The statistics of the datasets are shown in Table~\ref{tab:data-statistics}.

% \methodname achieves this by finding one representative dataset for each toxicity and extracting its keywords. In order to be more similar to the real world cases, all the dataset we used are collected from the real world Internet platform and with manually labeling.

Specifically, for abuse and hate speech, \methodname extracts keyword pairs from Social Bias Corpus~\cite{Sap2020SocialBF} and Dirty\footnote{https://github.com/pokemonchw/Dirty}.
Social Bias Corpus contains 150k structured annotations of social media posts, covering over 34k implications about a thousand demographic groups. Dirty is an open GitHub repository containing 2.5k Chinese toxic sentences with abusive and sexual words.

For malicious advertisement, \methodname extracts keyword pairs from SMS Spam Collection \footnote{https://www.kaggle.com/uciml/sms-spam-collection-dataset} and SpamMessage\footnote{https://github.com/hrwhisper/SpamMessage}. SMS Spam Collection is a set of tagged SMS messages, containing 5,574 SMS messages in English, tagged as being ham (legitimate) or spam. The data was manually extracted from the Grumbletext website, a UK forum in which cell phone users make public claims about SMS spam messages, while SpamMessage is an open GitHub repository containing 60k malicious advertisement messages.

For pornography, \methodname utilizes Sexting\footnote{https://github.com/mathigatti/sexting-dataset} and Midu \cite{Song2021EvidenceAN}, where Sexting is an English  pornographic text dataset containing 537 sexual texting messages, while Miduis a Chinese novel paragraph dataset collected by ourselves from an online literature reading platform called MiDu App\footnote{http://www.midureader.com/}. It is a corpus with 62,876 paragraphs including 7,360 pornographic paragraphs and 55,516 normal paragraphs.

\begin{table}
\centering
\caption{Statistics of toxic datasets.}
\begin{tabular}{l r r r r}
\toprule
\bf Dataset & \bf \#Sent & \bf Lang &  \bf Type & \bf Source\\
\midrule
Social Bias & 150k  &  English & Abuse\&Hate   &  Twitter \\
Dirty  &  2.5K & Chinese & Abuse\&Hate  & Weibo \\
SMSSpam & 5.5k & English &  Advertisement & Grumbletext \\
SpamMessage & 60K & Chinese & Advertisement & Taobao\\
Sexting  & 0.5K & English & Pornography & Github \\
Midu  & 7.3K & Chinese & Pornography & Midu \\
\bottomrule
\end{tabular}
\label{tab:data-statistics}
\end{table}

It is worth noting that not all words in the above datasets are potential keywords.
An ideal keyword should be frequently used in toxic content while less frequently in a general domain corpus so that it is more likely to contain toxicity.
Therefore, we use TF-IDF, a numerical statistic that reflects how important a word is to a document in a collection or corpus, to select potential keywords from the above datasets. The TF–IDF value increases proportionally to the number of times a word appears in the document and is offset by the number of documents in the corpus that contain the word.
Particularly, \methodname utilizes sklearn\footnote{https://scikit-learn.org/} (English) and Jieba library\footnote{https://github.com/fxsjy/jieba} (Chinese) to filter out the stop words, followed by calculating the TF-IDF score, and select the top $20$ words with the highest TF-IDF score as the candidate keywords for each dataset.

% 感觉可以放到threats
% Again, we admit that the keywords we extracted can be more diverse if we use more dataset or external knowledge base. But we just aim to show that \methodname can find a large number of failures of content moderation software and model even with the commonly used toxic words.

% \subsubsection{Slot Filling}
\subsubsection{Keyword Pair Extraction and Filtering}
% The final step of generating seed sentences is filling the appropriate keywords into the defined templates. 

After obtaining the candidate keywords, it is non-trivial to extract keyword pairs that can constitute toxic texts according to the templates, since not all keyword pairs are suitable.
For example, for the template of pornographic (\ie "A you/my B"), A should be a verb and B should be a sexual-related organ. It is inappropriate to fill a verb keyword into slot B, leading to meaningless sentences.

In order to find proper keyword pairs suitable for different templates, \methodname has to first obtain the property of every keyword. 
Considering that a keyword may have multiple properties, for each keyword, \methodname first retrieves 5 sentences containing the keyword from the corresponding dataset. 
Then, it utilizes the language analysis method to perform Part-of-Speech tagging (PoS tagging), which identifies the word property (\eg noun, verb, adjective or adverb), and Named Entity Recognition (NER), which determines whether the keyword is a belongs to a pre-defined category such as group names or location. 
The results of each keyword are voted on by the results of the five sentences.
In the implementation, \methodname adopts Flair toolkit\footnote{https://github.com/flairNLP/flair} for English and Baidu NLP API\footnote{https://ai.baidu.com/tech/nlp\_basic/} for Chinese. 

After obtaining the keyword property, \methodname extracts keyword pairs for three types of toxic contents as follows:

For abuse and hate speech, the template is "A is/are B", where A is a group and B is a negative noun or adjective. Hence, for A \methodname selects the keywords being identified as nouns by PoS tagging and group names by NER. And for B, \methodname selects the keyword which is either a noun or adjective and also has a negative sentiment, which is obtained from the APIs provided by Google\footnote{https://cloud.google.com/natural-language} (for English) and Baidu\footnote{https://ai.baidu.com/tech/nlp\_apply/sentiment\_classify} (for Chinese).
In addition, we require A and B to be different words, aiming to prevent the “monkey is monkey” situation that generates non-toxic seed sentences.

For malicious advertisement, the template is "A: B", where A is a product and B is the contact information. Hence, \methodname selects the keyword identified as the noun for A, according to the PoS tagging toolkit. For B, \methodname extracts the keywords with a prefix of a kind of contact way. We collect a candidate list, including "Tel", "Email", "WhatsApp" and "Ins". In this way, \methodname can extract keyword pairs that can constitute the prefix contact way followed by the specific contact information, such as "Tel: 12345678".

For pornography, the template is "A your/my B", where A is a verb and B is a sexual-related organ. Hence, \methodname extracts the keyword identified as verb for A and noun for B, according to the PoS tagging toolkit. 

% \subsubsection{Summary of seed sentences generation}
In this way, \methodname can extract keyword pairs as seeds for the further test case (\ie multimedia content) generation. Particularly, \methodname generates 100 keyword pairs for each type of toxicity and each language, ending up with 100*3*2=600 keyword pairs.

\subsection{Modality Transformation}
\label{sec-mrs:dist}
Since the extracted keyword pairs only contain single modality information (\ie textual), in order to generate multi-modal contents, \methodname has to perform modality transformation.
Particularly, this work focuses on three kinds of modalities, \ie visual, textual, and audio.
For each extracted keyword pair $(A, B)$, \methodname transforms the information of $A$ and $B$ to different modalities as follows:

\noindent \textbf{Visual modality.}
For the keyword pair $(A, B)$, \methodname transforms the textual information to visual modality via calling the Google figure API to search the top 5 images with $A$ or $B$ as the query. 
To ensure that the returned figures correctly contain the information of the queried keyword, \methodname further utilizes Baidu Image Recognition API\footnote{https://ai.baidu.com/ai-doc/IMAGERECOGNITION/Xk3bcxdum} to check the recognized salient object in each figure. 
If the object is not equal to the keyword, \methodname discards the figure.

\noindent \textbf{Textual modality.}
Since the keywords are texts {\em per se}, \methodname directly uses the templates for keyword extraction (Section~\ref{sec-mrs:pair-select}) to obtain the textual information.

\noindent \textbf{Audio modality.}
In order to transform the textual information to audio modality data (\ie speech), \methodname calls the Baidu text-to-speech synthesis API\footnote{https://ai.baidu.com/ai-doc/SPEECH/jk38y8gno}.

% For image test cases, the distributed modalities are vision and text. To convert textual information, i.e., A or B, to vision modality, we call google figure API to search the top 5 images with A/B as keywords. To ensure that the returned figure is A/B, we call Baidu Image Recognition api\footnote{https://ai.baidu.com/ai-doc/IMAGERECOGNITION/Xk3bcxdum} which can automatically return the name of the salient object in each figure. If the returned name is not equal to A/B, we discard this image.

% For video test cases, since a video can contain three modalities, \ie vision, audio and text, we need to convert some textual information to either vision or audio modality. For text to vision, the procedure is the same as the procedure of image test cases. For text to audio, we call Baidu text-to-speech synthesis api\footnote{https://ai.baidu.com/ai-doc/SPEECH/jk38y8gno} to generate speech.

\subsection{Multi-modal Semantic Fusion}
\label{sec-mrs:fuse}

After obtaining the information of keyword pairs from different modalities, \methodname performs semantic fusion, which fuses the multi-modal information together and generates test cases (\ie multimedia contents).
In particular, \methodname generates two kinds of multimedia data, \ie images and videos.
The images are generated by fusing visual and textual information, while the videos can be constituted by any combination of the three modalities, \ie visual \& textual, visual \& audio, audio \& text, and visual \& textual \& audio. Reminding that, besides the (A, B) keyword pair, there is a middle word in our template that represents the logical connection between A and B. For example. in abuse and hate speech, there is an "are" between A and B. We add this middle word to text or audio modality.

\subsubsection{Image Generation}

Image generation is performed via fusing visual and textual information (\ie image and text), more specifically, inserting the text into the image. 
In this process, we need to address three main issues: (1) how to decide the size of the text to avoid being too big or too small; (2) how to decide the location for the insertion, which should not affect the image content; (3) how to decide the color of the text so that it can be recognized in the image. 
In the following, we introduce three algorithms adopted by \methodname to address these issues, respectively.

\textbf{Text Resizing.} The goal of text resizing is to resize the text object to make the inserted text a comparable size with the salient object in the image. 
To achieve this target, \methodname first utilizes Baidu Image Recognition API to detect the salient object in the image and obtains the coordinates of the four vertices of its bounding box $(x_1,y_1), (x_1,y_2), (x_2,y_1), (x_2,y_2)$. 
Then, the height $h$ and the width $w$ of the salient object can be approximated as $w=x_2 - x_1$ and $h= y_2 - y_1$, respectively. 
Finally, \methodname sets the area of text within the range of $[0.8 * w * h, 1.2 * w * h]$, where 0.8 and 1.2 are hyper-parameters that we manually set based on empirical experiences.

\textbf{Location Determination.} After resizing the text, \methodname should select a suitable location in the image to insert the text. An ideal location should (1) have few overlaps with the salient object in the image, which is beneficial for humans to recognize both the text and the image; (2) the relative positions of the image and text follow the reading habits which is easier for human to understand the logic relation between the image information and the text information.

To find the location with few overlaps with the salient object, we first define 9 candidate insert positions: top left, top middle, top right, middle left, middle middle, middle right, bottom left, bottom middle and bottom right. 
Since \methodname has obtained the coordinates of the four vertices of the salient object in text resizing, it can directly calculate the overlapping area between the salient object and each candidate position.
If the overlapping area is larger than 30\%, \methodname discards this candidate position. 

To find the position in line with human reading habits, \methodname utilizes a rule-based method based on the consideration that humans typically read from top to bottom and from left to right.
It is worth noting that the keyword pair $(A, B)$ is extracted according to the templates (\ie "A is/are B", "A: B", and "A your/my B"). In all the used templates, a human reads $A$ first and reads $B$ later. 
Hence, object A should be above or to the left of object b, such constraints help \methodname filter out some candidate positions. 
For example, suppose we need to insert text $B$ into image $A$, the candidate positions of B following human reading habits are middle right, bottom middle and bottom right. For each (image, text) pair, if there is no ideal location candidate left based on the criteria above, \methodname discards this pair. If more than one candidates are suitable, \methodname randomly selects one as the location to insert the text.

\textbf{Font Color Selection.} After determining a location to insert the text, another issue \methodname needs to decide is the font color. If the font color is too similar to the background image color, the text will be hard for humans to recognize, leading to an invalid image.
To mitigate this issue, \methodname adopts a special webkit property called Text Stroke, which adds an exterior border around each character of the text. Text stroke can change the outline of the text, such as setting a color different from the original font color, so that the text can be recognized easily whatever the background color.

% \subsubsection{Fusing Image and Audio}
\subsubsection{Video Generation}

Unlike generating image data, video data can be generated by fusing any two or more information in different modalities.
\methodname conducts semantic fusion on different modality data as follows: 

\textbf{Fusing vision and text.} Generating video test cases that fuse vision ad text is similar to image generation, except that we should take the order of showing the image and the text into consideration. Similar to the position selection when generating images, for a keyword pair $(A, B)$, $A$ should come first and $B$ should come later. 
Hence \methodname generates the video that presents $A$ first and then presents $B$. For example, if $A$ is n image and $B$ is the text. The generated video shows $A$ first for a while and then shows the text of $B$. 
On the other hand, if $A$ is the text and $B$ is an image, the video would show the text of $A$ and then show the image $B$. In the implementation, \methodname adopts ffmpeg\footnote{https://ffmpeg.org/}, a complete, cross-platform API to record, convert and stream audio and video.

\textbf{Fusing vision and audio.} \methodname generates the video based on vision and audio by showing the image and playing the audio (a synthesized speech generated from the text). To make the video easier for humans to understand, again, we consider the order of showing the image and playing the audio. 
Specifically, for a keyword pair $(A, B)$, the generated video shows $A$ first and then shows $B$. For example, if $A$ is an image and $B$ is the audio. We show $A$ first for a while and then play the audio of $B$. On the other hand, if $A$ is the audio and $B$ is an image, the video first plays the audio of $A$ and meanwhile shows a blank video screen. The image $B$ will not be displayed until audio A has finished playing.

\textbf{Fusing text and audio.} Generating the video test cases with audio and text information is similar to generating the video that fuses image and audio. The main difference is that here \methodname shows the text, rather than the image, and plays audio. Again, if $A$ is audio and $B$ is text, the generated video plays the audio first and then shows the text, and {\em vise versa}. Since this kind of fusion contains both text and audio modality, we randomly add the middle word to either of the modality.

Finally, it is also feasible to generate videos by fusing all three modalities. To achieve this, \methodname adopts the procedure of fusing image and audio. The only difference is that here \methodname shows the middle words in text format between showing A and showing B, rather than showing accompanied with the keyword in audio modality.

As such, \methodname can generate multi-modal images and videos, which are used as test cases to detect errors in content moderation software.

\subsection{Toxicity Collection and Error Detection}
\label{sec-mrs:detection}

After modality transformation and semantic fusion, \methodname constructs test cases (\ie multi-modal images and videos).
Each test case has a corresponding keyword pair $(A, B)$, which is extracted based on a specific template. By filling the keyword pair into the template, we can get a {\em seed sentence}, which is further used for solving the test oracle problem.
During testing, \methodname first feeds the generated test cases, as well as the corresponding seed sentences, to the content moderation software under test. 
Since all seed sentences are supposed to be toxic, any test case whose corresponding seed sentence is classified as non-toxic by the textual content moderation software will be discarded. 

After toxicity collection, \methodname inspects the predicted toxicity of each test case. Any case that violates the metamorphic relation (MR) will be reported as a suspicious error. The MR is designed based on the following simple consideration: the toxicity should remain the same when some of the information is transformed into another modality. Since \methodname assumes seed sentences are all toxic ones, all the remaining cases should be classified as toxic, \ie an error is found if the test case is categorized as non-toxic.

In the implementation, we test 5 commercial software products provided by large Internet companies, \ie Google Cloud\footnote{https://cloud.google.com/video-intelligence/docs/analyze-safesearch}, Amazon Web Service\footnote{https://docs.aws.amazon.com/rekognition/latest/dg/moderation.html}, Baidu Cloud\footnote{https://cloud.baidu.com/doc/ANTIPORN/s/6ki012lqu}, Tencent Cloud\footnote{https://cloud.tencent.com/document/product/1235} and Alibaba Cloud\footnote{https://help.aliyun.com/document\_detail/146716.html}, all of which are the official content moderation software from big technology companies with more than 100 millions of users. In particular, all the software products are the latest version by Nov. 1st, 2022, when the experiments were conducted. The version information of the software under test is listed in Table~\ref{tab:software_version_info}.
Besides commercial software products, we also test popular research models. The core of multimedia content moderation software is image classification, hence we test Resnet-based image classification model\footnote{https://huggingface.co/microsoft/resnet-50} and Vision-Transformer-based image classification model\footnote{https://huggingface.co/google/vit-base-patch16-224}, both having more than 100k downloads according to Hugging Face model zoo, a famous AI model repository.

\begin{table}
\centering
\caption{Software Version Information.}
\begin{tabular}{l | r  r }
\toprule
\bf Software & \bf Version & \bf  Lanch Date   \\
\midrule
Google& builtin/stable & 2022.05.05 \\
Amazon&  5.0 & 2022.10.01 \\
Baidu &  4.16.3 & 2022.03.25 \\
Tencent&  2022-06-30 & 2022.06.30\\
Alibaba & 2022.06.15  & 2022.06.15  \\
\bottomrule
\end{tabular}
\label{tab:software_version_info}
\vspace{-12pt}
\end{table}

%In addition, \methodname also use widely-adopted textual content moderation to determine the toxicity of seed sentences
%Since Amazon Web Service dose not provide textual content moderation service, it also utilizes the Google Jigsaw’s Perspective\footnote{https://www.perspectiveapi.com/}.

%% file: Sections/4_Evaluation.tex
\section{Evaluation}
\label{sec-experiment}

To validate the effectiveness of \methodname and get more insights on enhancing content moderation software, we use our method to test five commercial software products and two state-of-the-art research models for multimedia content moderation. In this section, we detail the evaluation process and empirically explore the following four research questions (RQs).
\begin{itemize}[leftmargin=*]
    \item RQ1: Are the test cases generated by \methodname toxic and realistic?
    \item RQ2: Can \methodname find errors in content moderation software?
    \item RQ3: What factors affect the performance of \methodname?
    \item RQ4: Can we use \methodname to improve the performance of multimedia content moderation?
\end{itemize}

\subsection{RQ1: Are the test cases generated by \methodname be toxic and realistic?}

%\subsubsection{Seed Data Collection}

%\subsubsection{Validating the Generated Test Cases}

\methodname aims to generate test cases that are toxic and are as realistic as the ones real-world users produce to evade moderation.
Thus, in this section, we evaluate whether the generated test cases are still toxic (i.e., semantic-preserving) and whether they are realistic.

We conduct human annotation via crowd-sourcing. First, we generate 10 images and 30 videos (10 vision + text, 10 vision + audio, 10 audio + text) for each task and each language, ending up with 240 test cases for annotation.  For each test case, we ask three questions: 
 1) From "$1$ strongly disagree" to "$5$ strongly agree", to what extent do you agree that the image/video is semantically equivalent to the sentence? 
 2) From "$1$ strongly disagree" to "$5$ strongly agree", to what extent do you agree that the image/video is toxic (hate speech, malicious advertisement, or pornography, according to the dataset)? 
 3) From "$1$ strongly disagree" to "$5$ strongly agree", to what extent do you agree that this kind of image/video is realistic that Internet users would use?  
 For English/Chinese, we distribute the questionnaire and recruit 20/20 crowd workers on Prolific\footnote{https://www.prolific.co/}/Tencent Wenjuan\footnote{https://wj.qq.com/}, who have English/Chinese as their first language. Before annotation, we provide instructions about the type of questions and asked them to make subjective judgments in the annotation. We do not provide additional training to avoid potential bias from us. Then the annotators are asked to annotate the test cases.  Annotation results show that: 1) the generated test cases are semantically equivalent to the seed sentence, with an average score of 4.46/4.59; 2) the generated test cases are toxic, with an average score of 4.19/4.29; 3) the generated test cases are realistic, with the average  score of 3.96/4.15. We followed \cite{Kirk2021HatemojiAT} to measure the inter-annotator agreement using Randolph’s Kappa, obtaining a value of 0.84/0.81 for the test cases, which indicates "almost perfect agreement". There are a few cases that the generated images are annotated as non-toxic subjectively. For example, sometimes Google Image API returns a lovely cartoon image which could make the annotator feel less offensive.
 % how much do you regard this image/video is semantically equal to the sentence
 % how much do you regard this image/video is toxic
 % how much do you think this kind of image/video is realistic that Internet users would use

\begin{tcolorbox}[width=\linewidth, boxrule=0pt, colback=gray!20, colframe=gray!20]
\textbf{Answer to RQ1:}
Test cases generated by \methodname are toxic and commonly seen in real-world scenarios.
\end{tcolorbox}

\subsection{RQ2: Can \methodname find errors in content moderation software?}
\label{subsec-testing-software-with-tool}

\methodname aims to automatically generate test cases to find potential bugs in current content moderation software.
Hence, in this section, we evaluate the number of errors that \methodname can find in the outputs of commercial content moderation software and academic models.

\subsubsection{Software Products and Models under Test}

We use \methodname to test five commercial content moderation software products and two state-of-the-art research models. 

\paragraph{Image Content Moderation} For commercial software products, we choose the products that have been deployed in the cloud of large Internet companies, including Google, Amazon, Baidu, Tencent, and Alibaba. All of them can be accessed by registered users via an API. In particular, since Amazon Web Service does not provide advertisement detection services, Google Cloud does not provide hateful image detection and advertisement detection services, and Tencent Cloud does not provide hateful and abusive image detection services, we do not include them in our experiments. 
For research models, we choose the Vision-Transformer Model and ResNet-18 Model, which are the most downloaded image classification models in  Hugging Face\footnote{https://huggingface.co/} model zoo. For hate speech detection, we use the Hateful Memes Dataset~\cite{Kiela2020TheHM}, a hate and abuse detection dataset containing $8,500$ multimodal memes with labeled toxicity. For malicious advertisement detection, we use the Advertisement Understanding Dataset~\cite{Hussain2017AutomaticUO}, an advertisement dataset containing $64,832$ Ad and $13,597$ non-Ad images. Since there is no publicly available pornographic image detection dataset, we do not include this in our experiments.

We follow the official hyper-parameters setting to train the research model. More specifically,  we fine-tuned the model with $15$ epochs, a batch size of $8$, and a weight decay of $0.01$. We use the model with the highest accuracy on the validation set, and test its performance on the test set.

\paragraph{Video Content Moderation} All the five commercial software products mentioned above provide video content moderation services. Similar to the image content moderation experimental settings, we conduct experiments on the toxic content category if the APIs provide the corresponding detection services. Thus, there are no results on hate detection in Google Cloud and advertisement detection in Amazon Web Service, Google Cloud, and Tencent Cloud. Since the video content moderation service in Amazon Web Service and Google Cloud do not support audio modality analysis, we consider the modality fusion between vision and textual for these two APIs. Since there is no publicly available multi-modal video classification dataset for hate, pornographic, and advertisement detection, we do not conduct testing on the research model for video content moderation.

\begin{table}
\centering
\caption{Test case statistics.}
\begin{tabular}{l r | r r}
\toprule
\bf Software & \bf Tasks & \bf Seed Num  &\bf Toxic Num  \\
\midrule
\multirow{3}{*}{Google}& Abuse\&Hate & 100  &  81\\
 & pornography   & 100  &  74 \\
 & Advertisement  & 100 &  -\\
\midrule
\multirow{3}{*}{Baidu} & Abuse\&Hate & 100 & 89 \\
 & pornography &  100  &  95\\
& Advertisement & 100 & 91\\
\midrule
\multirow{3}{*}{Tencent}& Abuse\&Hate  & 100 &84 \\
& pornography & 100 &89 \\
& Advertisement  & 100 &91  \\
\midrule
\multirow{3}{*}{Alibaba}& Abuse\&Hate  & 100 &74 \\
& pornography & 100 &85 \\
& Advertisement  & 100 &84  \\
\bottomrule
\end{tabular}
\label{tab:test_case_stat}
\end{table}

For each tested software product or model, we first input all the seed data and filter out the sentence that cannot be identified as toxic. In other words, we only used those that have already been recognized as toxic contents as seed data to generate test cases. The statistics of seed data are shown in Table \ref{tab:test_case_stat}, which shows that most of the generated seed sentences are toxic. Then, we generate image and video test cases accordingly. Finally, we use the generated test cases to test the software products and research models. 

To evaluate how well \methodname does on generating test cases that trigger errors, we calculate Error Finding Rate (EFR), which is defined as follows:
$$\text{EFR} = \frac{\text{the number of misclassified test cases}}{\text{the number of generated test cases}}.$$

% \jz{感觉可以有一部分的分析放到discussion里，这里侧重EFR}
\subsubsection{Analysis}

We list the results in Table \ref{tab:abuse_image} and \ref{tab:abuse_video}. 

\begin{table*}
\centering
\caption{Error finding rates of image content moderation software and academic models (AM).}
\begin{tabular}{l  | c c c c c c c}
\toprule
\bf  &  \bf Amazon & \bf Google & \bf AM\_ResNet & \bf AM\_ViT  &  \bf Baidu  &\bf Tencent & \bf Alibaba \\
\cmidrule(lr){1-1} \cmidrule(lr){2-8} 
\bf Abuse Detection & 91.7\% & - & 94.1\% & 90.6\%& 99.55\% & -&76.50\%\\
\bf Porno Detection &  78.82\% & 77.18\% & - & - & 96.49\% & 76.77\% &33.14\%\\
\bf Ad Detection &  - & - & 78.8\% & 75.3\% & 99.78\% & 99.29\% &65.46\%\\
\bottomrule
\end{tabular}
\label{tab:abuse_image}
\end{table*}

\begin{table*}
\centering
\caption{Error finding rates of video content moderation software.}
\begin{tabular}{l  | l l l l | l l l l l | l l}
\toprule
\bf Perturbation   & \multicolumn{4}{c}{\bf Abuse Detection}   & \multicolumn{5}{c}{\bf Porn Detection}  & \multicolumn{2}{c}{\bf Ad Detection}\\
\cmidrule(lr){2-5} \cmidrule(lr){6-10}  \cmidrule(lr){11-12}  
\bf Modalities &  \bf Amazon & \bf Baidu  &\bf Tencent & \bf Alibaba  &  \bf Amozon & \bf Google  &\bf Baidu & \bf Tencent & \bf Alibaba & \bf Baidu  &\bf Alibaba\\
\midrule
\bf V + T & 89\% &71.91\% &100\% &60.81\% & 91.38\% & 100\% &50.53\% &100\% &60\% &0\%  &100\%\\
\bf V + A & - &100\% &100\% &54.05\% & - & - &91.58\% &100\% &55.29\% &5.49\%  &100\% \\
\bf A + T & - &32.58\% &100\% &81.08\% & - & - &10.53\% &100\% &58.82\% &0\%  &100\% \\
\bf V + A + T & - &76.67\% &100\% &97.29\% & - & - &55.79\% &100\% &97.64\% &7.61\% &100\%\\
\bottomrule
\end{tabular}
\label{tab:abuse_video}
\end{table*}

\paragraph{Overall Analysis}In general, \methodname is able to find errors in software products and research models with a relatively high EFR. Since most of these test cases are annotated as toxic according to our annotators, we believe such high EFR implies the effectiveness of \methodname and reveals the unexpected vulnerability of widely-deployed software products and research models.
%Figure~\ref{fig:example} shows images generated by \methodname that are labeled as non-toxic by existing detectors.

% \begin{figure*}
%      \centering
%      \begin{subfigure}[b]{0.3\textwidth}
%          \centering
%          \includegraphics[width=\textwidth]{Figures/hate.png}
%          \caption{Abuse\&Hate}
%          \label{fig:y equals x}
%      \end{subfigure}
%      \hfill
%      \begin{subfigure}[b]{0.3\textwidth}
%          \centering
%          \includegraphics[width=\textwidth]{Figures/porn.png}
%          \caption{Pornography}
%          \label{fig:three sin x}
%      \end{subfigure}
%      \hfill
%      \begin{subfigure}[b]{0.3\textwidth}
%          \centering
%          \includegraphics[width=\textwidth]{Figures/ad.png}
%          \caption{Advertisement}
%          \label{fig:five over x}
%      \end{subfigure}
%         \caption{Generated examples that are labeled as non-toxic by the content moderation software}
%         \label{fig:example}
% \end{figure*}

One common concern about AI software testing is whether the software performs well on existing test cases, which are toxic single-modal inputs in content moderation. To address this concern, we conduct a lightweight experiment to evaluate the effectiveness of the software under test in detecting toxic contents from the Internet. Since there is no publicly-available toxic (hateful, porno, and malicious ad) image benchmark dataset (probably due to the toxic nature), we manually collect a dataset with 50 hateful images, 50 porno images, and 50 ad images from the Internet. The average detection rate of five content moderation software is 97.8\%, indicating the effectiveness of the software. Thus, we think the high EFR achieved by \methodname is exciting.

\paragraph{The comparison between image and video} The robustness of image content moderation and that of video content moderation do not strictly hold a positive correlation. For example, Baidu's image moderation is much worse than its own video moderation, even though the testing logic should be similar. We think the algorithms behind them are different.

\paragraph{The comparison between different software} Since the seed sentence and test cases are different, it is difficult to compare the APIs for English content, \ie Amazon Web Service and Google Cloud, with the APIs for Chinese content, \ie Baidu Cloud, Tencent Cloud, and Alibaba Cloud. Even if the APIs for the same language are using the same test cases, it is also difficult to obtain an interesting conclusion because the functionalities of different software could vary a lot.

\paragraph{The comparison within a product}The performance of moderating different kinds of toxic data for one software product is different. For example, the robustness of pornography detection is much better than that of advertisement detection for the image content moderation APIs provided by Baidu, Tencent, and Alibaba. We think it is because the bulk of Internet Companies' revenue comes from advertising so it is common to share the link, image, or video of advertisement between the users on these platforms. In addition, in Chinese culture, advertisement is not as toxic as pornography. Therefore, these companies seem to pay less attention to advertisement detection than pornography detection when developing their content moderation software.

% \paragraph{The comparison between different modalities} For video content moderation, \methodname can distribute the information of a seed sentence into the combination of any two or three modalities.

%The final results give out two insights: (1) Commercial software are more robust, compared with research models. (2) Diverse tasks show different importance for various companies.

\begin{tcolorbox}[width=\linewidth, boxrule=0pt, colback=gray!20, colframe=gray!20]
\textbf{Answer to RQ2:}
\methodname can find substantial errors in both commercial software and research models. 
%Besides, almost all content moderation methods, either from the industry or academia, fail to detect some of toxic content.
\end{tcolorbox}

\subsection{RQ3: What factors affect the performance of \methodname?}

This section explores the impact of two factors on the \methodname's performance.
First, we studied the impact of the location of the image and text. In theory, a video should be identified as toxic content if it contains a toxic image or text, no matter where the image or text is located. However, we found that changing the location of the image and text can affect the prediction of content moderation software. More specifically, we randomly select 20 seed sentences for each kind of toxicity and generate two kinds of vision + text video test cases with different locations: 1) the image and the text are showing on the left of the screen;  2) the image and the text are showing on the right of the screen. The test cases are fed into the video content moderation software products. We found that on average 3.2\% of video test cases can bypass the moderation after we move the image and the text from the left to the right of the screen. This result implies that we can find the best location for toxic images and text if we want to find more cases that can bypass moderation.

Second, we studied the impact of the selection of modality. In theory, a toxic content should be identified as toxic no matter which modality it is. For example, an abusive sentence should be categorized as toxic no matter if it is in text format or being converted to audio format. However, we found that for multimedia content moderation software, the sensitivity of different modalities is not the same. To this end, we randomly select 20 toxic seed sentences for each kind of toxicity and generate two kinds of video test cases: 1) pure text video that shows the sentence from the beginning to the end; 2) pure audio video that uses text-to-speech APIs to convert the seed sentence to audio and playing the audio. The test cases are fed into the video content moderation software products. We found that on average 94.7\% of the pure text video can be detected as toxic while only 84.2\% of the pure audio video can be detected. This result implies that we can distribute more toxic information to the less sensitive modality if we want to find more cases that can bypass moderation.

\begin{tcolorbox}[width=\linewidth, boxrule=0pt, colback=gray!20, colframe=gray!20]
\textbf{Answer to RQ3:}
The location of the image and text and the modality can affect the performance of \methodname.
% We find two factors which can affect the performance of \methodname: (1)the location of image and text, and (2) the selection of modality.
\end{tcolorbox}

\subsection{RQ4: Can we use \methodname to improve the performance of multimedia content moderation?}

In this section, we discuss how to use \methodname to improve the robustness of multimedia content moderation. A natural way would be to retrain the models using the test cases generated by \methodname and check whether the retrained models become more robust. This method is known as the robust retraining method. For each research model, we randomly select 100 test cases that successfully bypass moderation and label them as toxic. After that, we add these 100 labeled images to the corresponding training set and re-train the ResNet-18 model and the Vision-Transformer model. We adopt the default fine-tuning settings from the Hugging Face website. In other words, the setting of robust training is identical to the setting of normal training. The only difference is the data.

To validate the effectiveness of robust retraining with \methodname, we use \methodname to test the model after robust retraining, denoted as "Rob", and compared its EFRs with the original model, denoted as "Ori". 
The results are presented in Table \ref{tab:abuse_improve}.
% We present the results of the model trained with the original dataset (denoted as Ori) and the model trained with robust retraining (denoted as Rob).
We can observe that robust training with \methodname's test cases largely reduces the EFRs on all the settings, which shows that \methodname can effectively  improve the robustness of multimedia content moderation models. In addition, we also evaluate their model accuracy on the original test set and the results show that in all settings, model accuracy remains the same after robust training.

\begin{tcolorbox}[width=\linewidth, boxrule=0pt, colback=gray!20, colframe=gray!20]
\textbf{Answer to RQ4:}
Test cases generated by \methodname can effectively improve the robustness of multimedia content moderation models without affecting their accuracy.
\end{tcolorbox}

\begin{table}
\centering
\caption{Improvement of models with \methodname}
\centering
\begin{tabular}{l l l | l }
\toprule
\bf Model & \bf Task & \bf Ori  &\bf Rob  \\
\midrule
\multirow{2}{*}{ResNet-18}& Abuse & 94.1\%  & 5.7\% \\
& Ad &  78.8\% & 3.6\% \\
\midrule
\multirow{2}{*}{ViT}& Abuse & 90.6\% & 4.8\%  \\
& Ad & 75.3\% & 2.5\%\\
\bottomrule
\end{tabular}
\label{tab:abuse_improve}
\end{table}

%% file: Sections/5_Discussion.tex
\section{Discussion}
\label{sec-discuss}
%\sq{Maybe we can add some discussions on the FPs and FNs of our method in this section or in the evaluation section.}

% \jz{可以总结一下finding和insights，e.g.,现有工具的缺陷，哪些modality更有效，现有工具如何提升}
\subsection{Summary of Findings}

% 吹一波全面性和泛化性（三种modality, black-box）
\subsubsection{Different modalities matter.}
In the evaluation, we can observe that different content moderation software has different specialties. For example, the content moderation software of Baidu performs quite well (\ie low EFR) on videos fused by audio and text, while cannot achieve satisfying performance on those fused by vision and audio. In addition, the software from Alibaba can achieve 65.46\% EFR on image-based Ad detection, while it is infeasible to detect video-based advertisements (\ie 100\% EFR for all kinds of fused videos). Such characteristics make it non-trivial to comprehensively test content moderation software, \eg test cases for software may be inappropriate for another. 
In contrast, \methodname considers three different modalities, and performs semantic fusion to generate both image and video test cases in English and Chinese. Such design makes \methodname produce comprehensive test cases for different content moderation tasks.
In addition, \methodname is designed for black-box testing, which does not require white-box access to the software under test. Such design makes it convenient for adapting \methodname to other AI-based software testing tasks, such as image caption systems.

\subsubsection{Future Direction}

Although \methodname is able to find numerous errors in multimedia content moderation software and research models, there are several limitations that can lead to future works.

First, we can improve the diversity of our generated test cases. In this work, we only design one simple template and extract limited keywords from a toxic dataset for each kind of toxicity to generate test cases. In future work, we can follow the framework of this paper but use more kinds of templates and keywords.

Second, we can improve the authenticity of our generated test cases. In this work, we conduct multi-modal fusion by inserting text into an image or showing an image while playing a synthesized speech. In future work, we can utilize more advanced conditional image or video generation models to generate more vivid test cases.

% \jz{这一小节感觉可以删掉，泛化性可以放到summary of findings或者threats里面提}
% \subsection{Why \methodname}
% \begin{itemize}[leftmargin=*]
%     \item \textbf{Novelty}: \methodname ~is the first SE work to systematically test the reliability of multi-modal information fusion for multimedia content moderation software. Existing efforts generally focus on single modality. 
%     \item \textbf{Comprehensiveness} \methodname ~ considers different kinds of information fusion to generate both image and video tests cases for three kinds of toxic data and can test the reliability of industry software and academia models in both English and Chinese. 
%     \item \textbf{Generalization}: \methodname ~does not require white-box access to the tested model and can be easily extended to test other software products and research models, like image caption system.
% \end{itemize}

\subsection{Simplicity of the Methodology}
While our \textit{Semantic Fusion} methodology is simple, it is novel, general, and effective. The idea that fuses the semantics of single-modal inputs into a new input that combines the semantics of its ancestors has not been explored by any existing studies. Meanwhile, although the templates we use are simple, the single-model inputs can be automatically extracted from media sources or datasets to generate diverse test cases.  It can be easily generalized to more complicated templates and seeds. In addition, the test cases generated from our templates can already reveal many errors in real-world software, which was reported to be highly effective by the companies. We believe that our simple methodology could benefit the community and we hope our work will be a starting point that brings more attention to multimedia content moderation in the field.

\subsection{Threats to Validity}

\textbf{Internal Validity.} The major threat to internal validity is that the image and video test cases generated by \methodname may no longer preserve the toxic nature of the seed sentence, which may cause false positives during testing. Besides, \methodname only uses simple templates to extract keyword pairs for test case generation. 
The template design can also affect the quality of the generated test cases, which may not be realistic enough that seems to be applied by users and appear in the real world. To relieve these issues, we validated the generated test cases by conducting human annotations, which shows that our generated text cases are toxic and realistic. During the manual inspection of the evaluation results, we also observe few false positive.
In addition, \methodname adopts existing tools for data transformation, \eg Flair for PoS tagging and ffmpeg for video processing, whose performance may affect the quality of generated cases. To relieve this issue, we only chose tools that are widely used by millions of users. Moreover, these tools are also replaceable, which means users of \methodname can customize these tools for specific testing tasks.

\textbf{External Validity.} 
The major threat to external validity is whether the content moderation software products we test are good and representative of what the industry uses. To relieve this issue, we evaluated the effectiveness of these products. The average detection rate of five content moderation software products is 97.8\%, indicating their effectiveness. As for representativeness, all five commercial software products are paid cloud services provided by the could platform from the big companies. Moderation services and other cloud services have become an important source of income for them. Meanwhile, a huge amount of downstream companies and users are using paid services provided by these companies. The customer list can be seen from Google\footnote{https://cloud.google.com/customers\#/products=Data\_Analytics}, Amazon\footnote{https://aws.amazon.com/machine-learning/customers/} and Baidu\footnote{
https://cloud.baidu.com/partner/plan.html\#search}. Thus, we believe the moderation software studied in our paper has a significant impact on real users and is representative of industry practice.
The other possible threat to external validity lies in the source datasets used for keyword extraction. Low-quality keywords can directly affect the testing effectiveness of generated test cases. To relieve this threat, we extracted keywords from six practical datasets, which consist of real-world toxic contents.
Manual inspection and the evaluation results also prove that \methodname can produce effective multimedia contents.
Another threat may lie in the
comprehensiveness and representation of our metamorphic relations, which might hurt the generalizability of our results and findings. 
To mitigate this threat, we extensively implemented splitting and fusion for all the possible permutations of different modalities. Hence, the final metamorphic relations are empirically comprehensive. To further validate the generalizability of \methodname, we also conducted extensive experiments. Specifically, we tested \methodname on five commercial textual software products and two research methods for content moderation. 
We also tested \methodname with different kinds of toxic data: abuse and hate speech, porn content and malicious advertisement, in both English and Chinese. The evaluation results confirm that our findings can generalize to different methods and tasks.

%% file: Sections/6_Related_new.tex
\section{Related Work}
\label{sec_related}
% \jz{这一段内容有点多，精简一下}
\subsection{Content Moderation}

There are generally two categories of research models for content moderation, \ie feature engineering-based models and neural networks-based models. Feature engineering-based models work by manually designing features and using the features to train classification models to identify toxic contents. While a neural networks-based method is typically a neural network model trained in an end-to-end manner on a huge amount of data, which does not require much human effort on feature engineering.

\subsubsection{Single-modal Content Moderation}
In the literature, there are mainly three categories of data modalities, \ie visual, audio, and textual.

 \textbf{Visual Content Moderation}.
Early works on toxic image detection mainly extract some pre-designed features and then train a machine learning classifier. For example, Shen \etal ~\cite{Shen2010API} computed color histograms to detect pornography images, depending on the activity suspicion that pixels in pornographic images are mostly skin. Zhang \etal~\cite{Zhang2005SemisupervisedAH} extract the motion features and use a Hidden Markov Model (HMM) for video anomaly detection.
With the development of deep learning techniques and the building of big datasets, the progress of image representation learning has motivated researchers to explore neural network-based models for visual content moderation. For example, Moustafa~\cite{Moustafa2015ApplyingDL} adapted a CNN architecture for image classification to the pornographic video classification task. Each frame is input for being classified as porn or non-porn and then integrating the final result for a video via a majority voting process.

 \textbf{Audio Content Moderation}. Zhang~\cite{Zhang2005SemisupervisedAH} extracted Mel-Frequency Cepstral Coefficients (MFCCs) feature and use an HMM to detect audio events. \cite{Shi2009ObjectionableAC} extract various audio features, including MFCC, mean of short-time zero-crossing rates(mSTZCR), mean of the spectral centroid(
SC) and high short-time zero-crossing rates
ratio(HZCRR), and use in-class clustering for porno-sounds detection.
Gupta~\cite{Gupta2022ADIMAAD} utilized a bidirectional Gated Recurrent Units(GRU) and Long Short-Term Memory (LSTM) model to  detect abuse from multilingual audio. Recently researchers proposed a deep learning method that can detect COVID-19 from breath and cough audio~\cite{Coppock2021EndtoendCN}.

 \textbf{Textual Content Moderation}. Early works are rule-based methods that are based on pre-defined rules or dictionaries\cite{Spertus1997SmokeyAR,Razavi2010AAI}. However, rule-based can hardly deal with implicit abuse and sarcasm. Besides, they are vulnerable to errors in spelling, punctuation, and grammar \cite{Wiegand2018InducingAL}.  Computation-based methods leverage some statistics of the textual data, such as TF-IDF\cite{Yin2009DetectionOH,Salminen2018AnatomyOO}. Computation-based methods require less human effort and are more robust to spelling, punctuation, and grammatical variations. Nevertheless, this kind of approach can only capture surface-level patterns, not deeper semantic properties~\cite{Wiegand2018InducingAL}. The advancements in text representation learning have motivated researchers to explore neural network-based models, such as Feed-forward neural networks\cite{Djuric2015HateSD}, LSTM\cite{Badjatiya2017DeepLF} or the pre-training models~\cite{Devlin2019BERTPO,Liu2019RoBERTaAR}, on textural content moderation task and have achieved remarkable performance.

\subsubsection{Testing Multi-modal Content Moderation}

The testing of multi-modal content moderation is relatively unexplored compared with that of single-modal content moderation. Kiela et al.~\cite{Kiela2020TheHM} manually created and labeled a multi-modal hateful memes dataset as a benchmark to evaluate the multi-modal understanding ability of cross-vision and text modality. Hussain et al.~\cite{Hussain2017AutomaticUO} collected and labeled a TV advertisement classification benchmark that predicts whether a frame of video is an advertisement. \methodname is different from the existing papers: 1) existing work needs huge human effort to create and label the multi-modal test cases while \methodname can automatically generate and label the test cases; 2) existing work only provides image test cases on English hateful and advertisement image detection, while \methodname is able to comprehensively generate image and video test cases in two languages for three kinds of toxicity detection. In addition, our experiments show that \methodname can find numerous errors from models trained on the datasets provided by existing work. We believe our work complements this line of existing papers.

\subsection{Robustness of AI Software}

Artificial Intelligence (AI) software has been widely adopted by many domains, such as autonomous driving and face recognition. However, AI software is not robust enough and can generate erroneous outputs that lead to fatal accidents~\cite{notrobustself-driving,notrobusttesla}. To this end, researchers have proposed a variety of methods to generate adversarial examples or test cases that can fool AI software~\cite{Carlini2016HiddenVC,Tu2021ExploringAR, Luo2021InteractivePF,Wang2021PrioritizingTI,Shen2022test,Rozire2022LeveragingAU,Sun2020AutomaticTA,Zhang2022MachineLT,Gao2022AdaptiveTS,Liu2022DeepStateST,Ji2021AutomatedTF,Liu2021DialTestAT,Kim2021MultimodalSA,Kang2020SINVADSI,Kim2018GeneratingTI,Chen2020PracticalAE,Wei2022FreeLF,Tian2020TestingDI,Tian2018DeepTestAT, Zhang2023ImprovingTT, Zhang2022ImprovingAT}. Meanwhile, researchers have also designed approaches to improve the robustness of AI software, \eg the robust training mechanism~\cite{Madry2018TowardsDL}. 

AI software has also become adept at solving many NLP tasks, such as sentiment analysis~\cite{Zhang2017SentimentAA}, reading comprehension~\cite{Yu2018QANetCL}, Grammatical Error Correction~\cite{Wu2023ChatGPTOG} and machine translation~\cite{Bahdanau2015NeuralMT, Jiao2023IsCA, Jiao2022TencentsMM}. 
In recent years, inspired by the work on adversarial examples in the computer vision field, NLP researchers have started exploring attack and defense techniques for various NLP software~\cite{Gupta2020MachineTT, He2021TestingMT,Huang2022AEONAM}. For example, Ribeiro \etal~\cite{Ribeiro2020BeyondAB} designed a behavioral testing method to test NLP software for sentiment analysis, duplicate question answering, and machine comprehension. Li \etal~\cite{Li2020BERTATTACKAA} used deep learning models to generate test cases for deep learning-based NLP software. And Wang \etal~\cite{Wang2023MTTMMT} proposed a metamorphic testing framework for textual content Moderation software.
Unlike these studies, this work focuses on the robustness of multimedia content moderation software, which has not been explored by existing work.

% 这一段感觉可以直接放到related work
\subsection{Metamorphic Testing}

This work adopts metamorphic testing~\cite{Chen2020MetamorphicTA}, a widely-adopted method to solve the test oracle problem.
Specifically, it solves the test oracle problem via metamorphic relations, which describe the relations between bug-free software’s outputs on an input sample and that transformed by a pre-defined rule. Therefore, given an input sample, metamorphic testing transforms it into a new test case via a pre-defined transformation rule. Then by checking whether the outputs of the software on this pair of input samples satisfy the expected relation, metamorphic testing can identify the software bugs.

% \jz{这一段可以删掉，感觉太基础了}
%Consider a famous example that a program computing $\sin x$. According to mathematical property, a metamorphic relation for sine functions is "$\sin (\pi-x) = \sin x$". Thus, even though the expected value of $\sin x_1$ for the source test case is not known, a follow-up test case $x_2 = \pi-x_1$ can be constructed. We can verify whether $\sin x_1 = \sin x_2$ without knowing the concrete values of either sine calculation. Any inconsistency indicates a failure~\cite{Segura2016ASO}.

Metamorphic testing has been recently used for testing machine learning software. These efforts mainly focus on defining modifications in the dataset that can generate test cases to verify the quality of the model under test. Chen \etal~\cite{Chen2008AnIA} investigated the use of MT in bioinformatics applications. Xie \etal~\cite{Xie2011TestingAV} defined 11 MRs to test k-Nearest Neighbors and Naive Bayes algorithms. Dwarakanath \etal~\cite{Dwarakanath2018IdentifyingIB} presented 8 MRs to test SVM-based and ResNet-based image classifiers. Zhang \etal~\cite{Zhang2018DeepRoadGM} tested image-based autonomous driving systems by applying GANs to produce driving scenes with various weather conditions and checking the consistency of the system outputs.

%% file: Sections/7_Conclusion.tex
\section{Conclusion}

In this paper, we propose a general, effective methodology, \textit{Semantic Fusion}, for validating multimedia content moderation software and further realize it as a practical tool \methodname. \methodname can generate image and video test cases covering three kinds of toxic contents, including abusive language and hate speech, malicious advertisements, and pornography, in both two language settings (English and Chinese). We used \methodname to test five commercial content moderation software products and two research models. Results show that \methodname can find numerous errors from all the software and models. In addition, we showed that we can improve the models' robustness by utilizing the test cases produced by \methodname to perform robust training. We hope our framework can help identify the defects of current content moderation software and facilitate their development, contributing to a cleaner and better Internet environment.

\section{Acknowledgement}
The work described in this paper was supported by the Research Grants Council of the Hong Kong Special Administrative Region, China (No. CUHK 14206921 of the General Research Fund) and the National Natural Science Foundation of China (Grant Nos. 62102340 and 62206318)